\documentclass[twocolumn]{jpsj3}
\usepackage{txfonts}

\title{Measurement of Energy Relaxation in Quantum Hall Edge States Utilizing Quantum Point Contacts}

\author{\name{Tomohiro \surname{Otsuka}}$^{1,2}$\thanks{E-mail: tomohiro.otsuka@riken.jp}\thanks{These authors contributed equally to this work.}, 
\name{Yuuki \surname{Sugihara}}$^{2,\dagger}$, 
\name{Jun \surname{Yoneda}}$^2$, 
\name{Takashi \surname{Nakajima}}$^{1,2}$, 
and \name{Seigo \surname{Tarucha}}$^{1,2,3,4}$}

\inst{$^1$Center for Emergent Matter Science, RIKEN, 2-1 Hirosawa, Wako, Saitama 351-0198, Japan \\
$^2$Department of Applied Physics, University of Tokyo, Bunkyo-ku, Tokyo 113-8656, Japan \\
$^3$Quantum-Phase Electronics Center, University of Tokyo, Bunkyo-ku, Tokyo 113-8656, Japan \\
$^4$Institute for Nano Quantum Information Electronics, University of Tokyo, 4-6-1 Komaba, Meguro, Tokyo 153-8505, Japan} 

\abst{We demonstrated a method to probe local electronic states and energy relaxation in quantum Hall edge states utilizing quantum point contacts.
We evaluated relaxation lengths in two cases; first, with electron tunneling, and second, only with energy exchange without tunneling between edge states.
The results were consistent with previous experiments and validated our method with quantum point contacts.
We applied this method to measure the energy relaxation length around a hotspot in quantum Hall regimes and revealed the possible short-distance relaxation process to be the relaxation due to energy exchange between edge states without electron tunneling.}

\kword{quantum Hall effect, edge state, quantum point contact, local probe, energy relaxation, non-equilibrium distribution}

\begin{document}
\maketitle

\section{Introduction}

Edge states in the quantum Hall regime~\cite{1980KlitzingPRL} attract strong interests in basic science and applications to quantum electronics in semiconductor micro devices. 
Quantum Hall edge states have long coherence lengths in solids and well defined chirality.~\cite{1982HalperinPRB, 1988ButtikerPRB}
There is a good analogy between electrons propagating ballistically in edge states and photons propagating in optical media.
By utilizing these properties, interesting electronic devices, e.g. electronic Fabry-P\'erot interferometers,~\cite{1991BeenakkerSSP} Mach-Zender interferometers~\cite{2003JiNat, 2007NederNatPhys} and on-demand single electron sources,~\cite{2007FeveSci} etc. have been demonstrated and used to explore the quantum nature of the electronic states in solids.
Also, there are theoretical proposals to use quantum Hall edge states for quantum information processing applications by using the quantum states themselves or forming quantum networks connecting isolated quantum states like quantum dots.~\cite{2001IonicioiuIJMPB, 2004StacePRL}

In many of the applications, local electronic properties of the edge states and the spatial changes of the states induced by energy relaxation are important.
Recently, experiments probing the local electronic states, non-equilibrium energy distribution and energy relaxation in quantum Hall edge states utilizing quantum dots as local probes have been reported.~\cite{2010AltimirasNatPhys, 2010OtsukaPRB, 2010leSueurPRL, 2010AltimirasPRL}
The electronic energy distribution can be accessed by using the discrete energy levels formed in quantum dots as energy filters.
By measuring the spatial change of the electronic energy distribution, energy relaxation with a short relaxation length (3~$\mu $m), which originates from energy exchange between edge channels, was observed and the detail of the relaxation mechanism was discussed using a microscopic model.~\cite{2010leSueurPRL, 2010LundePRB}
A similar approach has also been used to probe the electronic states in the fractional quantum Hall regime to reveal charge and heat transfer.~\cite{2012VenkatachalamNatPhys}

In this paper, we measured the local electronic states in quantum Hall edge states using a different kind of probe: quantum point contacts (QPCs).~\cite{1988vanWeesPRL, 1988WharamJPhysC}
The QPC probes are easier to fabricate and can be applied to a wider range of experiments such as measurements in relatively high temperature conditions compared to the quantum dot probes.
With QPC probes, we were able to access the local electrochemical potential and also maybe the local electron temperature as we will show in this paper.
We used the QPC probes to investigate the spatial change of the local electronic states and evaluated the energy relaxation lengths.

In this paper first we applied the method to evaluate the relaxation length in the cases with electron tunneling and only with energy exchange without tunneling between edge states.
By comparing these results with previous experiments, we checked the validity of our method.
Next, we applied this method to probe energy relaxation around a specific energy dissipation point, called a ``hotspot'' in quantum Hall regimes, which is formed near a gate by applying a large source-drain bias across the gate to create hot electrons.~\cite{1992KomiyamaPRB}
We revealed the possible mechanism of the short-distance energy relaxation around the hotspot from the obtained relaxation length.

\section{Device and Measurement Scheme}

\begin{figure}[b]
\begin{center}
  \includegraphics{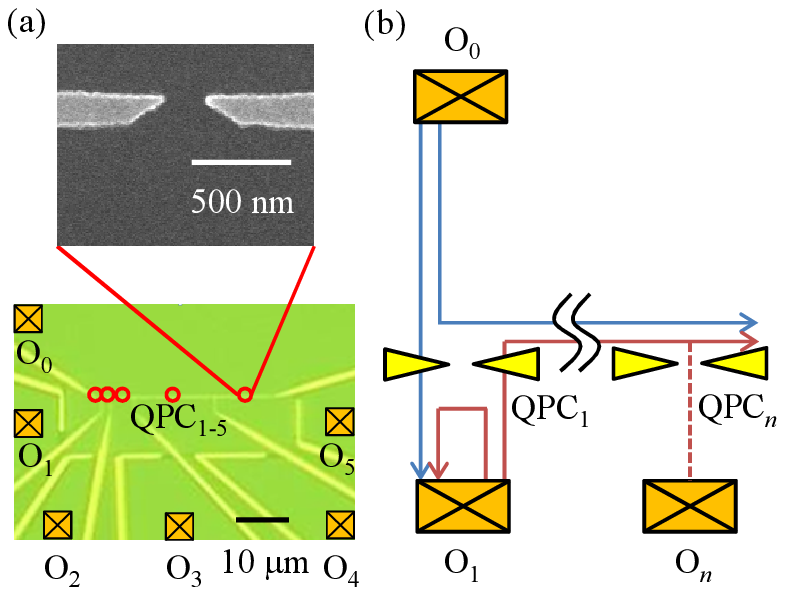}
  \caption{(color online) (a) Optical and scanning electron micrographs of the device.
  We prepared five QPCs to initialize and probe the local electronic states in the quantum Hall edge states.
  The distance from QPC$_1$ to QPC$_2$ - QPC$_5$ is 1.0, 3.1, 9.7 and 30.0~$\mu $m, respectively.
  (b) Schematic of the device operation. 
  We used QPC$_1$ to create a non-equilibrium electron distribution and QPC$_n (n=2,3,4,5)$ to probe the local electronic states of the outermost edge channel.
  The applied voltages on O$_0$ and O$_1$, $V_0$ and $V_1$, are defined independently.
  }
  \label{Device}
\end{center}
\end{figure}

Figure~\ref{Device}(a) shows optical and scanning electron micrographs of the device.
The device was fabricated from a GaAs/AlGaAs heterostructure wafer with sheet carrier density 3.9~$\times$~10$^{15}$~m$^{-2}$ and mobility 52~m$^2$/Vs at 2~K.
The two-dimensional electron gas is formed 60~nm beneath the surface.
We patterned a mesa structure by wet-etching and formed Ti/Au Schottky surface gates to define QPCs by metal deposition.
Five QPCs (QPC$_{1}$ - QPC$_{5}$) were formed by applying negative voltages
on the gates and each QPC was connected to each corresponding Ohmic contact O$_{1}$ - O$_{5}$.
The distance from QPC$_1$ to QPC$_2$ - QPC$_5$, $d$, is 1.0, 3.1, 9.7 and 30.0~$\mu $m, respectively.

We applied a magnetic field perpendicular to the wafer surface and created quantum Hall edge states in the mesa.
With the magnetic field of 4.05~T, quantum Hall states with filling factor $\nu $=4 were formed in the bulk.
At this magnetic field, the spin splitting was not large enough to create fully spin split edge channels in our device at 2~K,
so spin-degenerate two edge channels were formed in the device. 
The outer channel consists of $\nu $=1 and 2, and the inner channel of $\nu $=3 and 4.
We injected electrons from the Ohmic contacts O$_0$ and O$_1$ to the edge states, and created a non-equilibrium energy distribution by changing the transmission of QPC$_1$ and the voltages applied on O$_0$, $V_0$, and O$_1$, $V_1$, respectively (Fig.~\ref{Device}(b)).
In the following experiments, we fixed $V_0=0$~V and changed only $V_1$.
We monitored the change of the local electronic states by measuring voltages at the probe Ohmic contacts O$_2$ to O$_5$.
By adjusting the conductance of one of the QPCs, QPC$_n (n=2,3,4,5)$, to 2$e^2/h$ and the others to 0, the outermost edge channel can interact with only one Ohmic contact O$_n$.
The voltage measured at O$_n$ using a voltage meter with high input impedance, $V_n$, reflects the local electronic states of the outermost edge channel at the position of QPC$_n$.
We can measure the change in the local electronic states as a function of $d$ and get information about the energy relaxation by measuring $V_n$ for the different probes QPC$_n$.

\section{Energy Relaxation with Electron Tunneling}

\begin{figure}[b]
\begin{center}
  \includegraphics{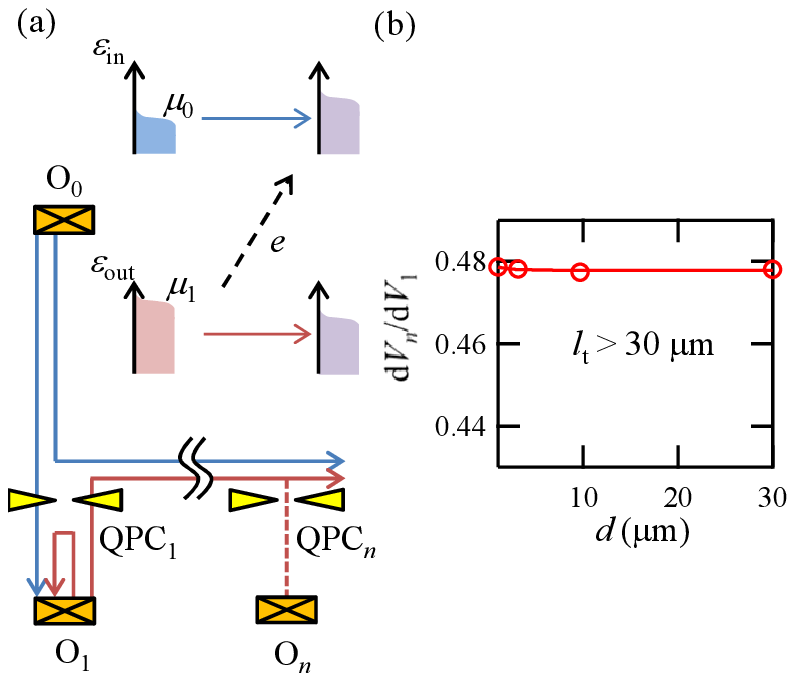}
  \caption{(color online) (a) Schematic of the experimental setup to probe the energy relaxation with electron tunneling. 
  The conductance of QPC$_1$ is set to $2e^2/h$ and the inner edge channel is reflected at QPC$_1$.
  The inner edge channel is connected to O$_0$ and the outer channel is connected to O$_1$.
  We can induce a difference in $\mu $ between edge channels by controlling $V_0$ and $V_1$, which causes the energy relaxation with electron tunneling.
  (b) ${\rm d}V_n/{\rm d} V_1$ as a function of $d$.
  No significant relaxation is observed up to $d=30~\mu$m.
  }
  \label{Tunnel}
\end{center}
\end{figure}

Energy relaxation in quantum Hall edge states can be classified mainly into two kinds of mechanisms.
The first one is tunneling of electrons between edge states.~\cite{1992MullerPRB, 1992KomiyamaPRB}
Assume edge states which have different electrochemical potentials $\mu $.
If high-energy electrons in the edge states with higher $\mu $ tunnel to the edge states with lower $\mu $, energy relaxation occurs to make $\mu $ of the edge states equivalent.
The other is energy exchange between edge states (and probably also between edge states and bulk states) without tunneling of electrons.~\cite{2010leSueurPRL} 
There is always Coulomb interaction between the edge states and this enables energy exchange between them even without tunneling of electrons.
For example, in the case of the edge states with different electron temperatures, the hot edge channel can radiate energy and then the energy relaxation, which equalizes electron temperature of the edge states, occurs by this process.

First, we measured the relaxation length of energy relaxation with the electron tunneling.
To create a non-equilibrium energy distribution between the edge states with different $\mu $, which causes the energy relaxation with electron tunneling, we connect the inner edge channel to contact O$_0$ and the outer edge channel to contact O$_1$ by adjusting the conductance through QPC$_1$ to $2e^2/h$ and reflecting the inner edge channel at QPC$_1$.
$V_0$ and $V_1$ determine the electrochemical potential of electrons supplied by O$_0$, $\mu _0$, and by O$_1$, $\mu _1$, respectively.
We can create a difference in $\mu $ between the edge states and realize $\mu _0<\mu _1$, by making $V_0>V_1$ (Fig.~\ref{Tunnel}(a)).
In this case, electrons in the outer edge channel with higher energy can tunnel into the inner channel contacting O$_0$.
This accompanies the energy relaxation with electron tunneling.
We can measure the relaxation length of this process $l_{\rm t} $ by monitoring $\mu $ of the outer channel through the measurement of $V_n $ at QPC$_{2}$ - QPC$_{5}$.

Figure~\ref{Tunnel}(b) shows the measured $V_n$ as a function of $d$.
To normalize the value of $V_n$ as $V_1$, we plot the numerical derivative of $V_n$, ${\rm d}V_n/{\rm d}V_1$.
In the low bias regime of $V_1$ (less than 100$~\mu$V), $V_n$ changes linearly with $V_1$ and ${\rm d}V_n/{\rm d}V_1$ can be defined uniquely.
If there is energy relaxation in this range of $d$, $V_n$ will decrease with increasing $d$ and we will observe the decrease of ${\rm d}V_n/{\rm d} V_1$.
But in Fig.~\ref{Tunnel}(b), the value of ${\rm d}V_n/{\rm d}V_1$ is always around 0.48 and we observe no such decrease of ${\rm d}V_n/{\rm d} V_1$ up to $d=30~\mu$m.
This result indicates $l_{\rm t}$ is much longer than $30~\mu$m, the maximum length accessible in this device geometry.~\cite{comment1}
This long $l_{\rm t}$, induced by the tunneling of electrons between the edge states consisting of well separated Landau levels $\nu = 2$ and 4, is consistent with the previous experiments using locally gated Hall bars with longer probe distances showing the long energy relaxation lengths (over 100~$\mu $m).~\cite{1992MullerPRB, 1992KomiyamaPRB, 1995HiraiPRB}

\section{Energy Relaxation Only with Energy Exchange}

\begin{figure}[b]
\begin{center}
  \includegraphics{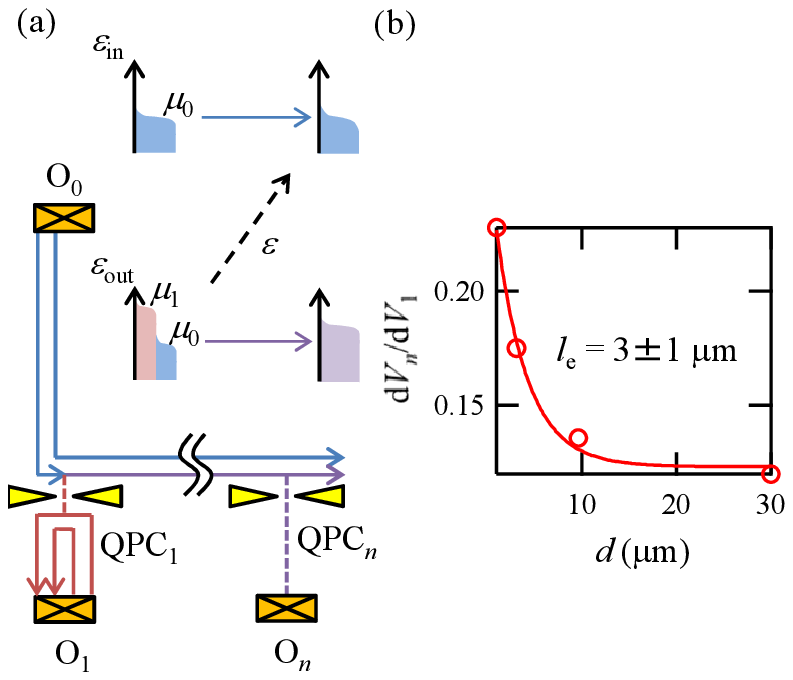}
  \caption{(color online) (a) Schematic of the experimental setup to probe the energy relaxation with energy exchange. 
  The conductance of QPC$_1$ is set to $e^2/h$.
  The inner edge channel is connected to O$_0$ and the outer channel is connected to both O$_0$ and O$_1$.
  The half of electrons in the outer edge channel is supplied by O$_1$ and the remaining half by O$_0$.
  We can create the two-step non-equilibrium energy distribution with two Fermi energies $\mu _0$ and $\mu _1$ in the outer edge channel by controlling $V_0$ and $V_1$.
  (b) ${\rm d}V_n/{\rm d} V_1$ as a function of $d$.
  Decay of ${\rm d}V_n/{\rm d} V_1$ is observed with the increase of $d$.
  The solid line is the result of fitting and the evaluated relaxation length $l_{\rm e}$ is $3\pm 1~\mu$m.
  }
  \label{Energy}
\end{center}
\end{figure}

Next, we measured the energy relaxation without electron tunneling and only with energy exchange between edge channels.
To induce such kind of relaxation, we set the conductance through QPC$_1$ to $e^2/h$.
The inner edge channel is connected to O$_0$ and the outer channel is connected to both O$_0$ and O$_1$.
The half of the electrons in the outer edge channel is supplied by O$_0$ and the remaining half by O$_1$.
This results in the two-step non-equilibrium energy distribution with two Fermi levels $\mu _0$ and $\mu _1$ in the outer edge channel when we make a difference between $\mu _0$ and $\mu _1$ (Fig.~\ref{Energy}(a)).~\cite{2010AltimirasNatPhys, 2010leSueurPRL}
In this case, the energy can be transferred from the outer to the inner edge channel because the outer channel can radiate energy in the process of the relaxation from the two-step non-equilibrium energy distribution to the conventional Fermi distribution.
We will measure this energy relaxation length $l_{\rm e} $ by monitoring $V_n $ of the outer channel at QPC$_{2}$ - QPC$_{5}$.
Note that the energy relaxation with electron tunneling is negligible on this short scale as we observed in Fig.~\ref{Tunnel}.
We can get the pure relaxation length only by energy exchange between the edge states.

Figure~\ref{Energy}(b) shows the measured ${\rm d}V_n/{\rm d} V_1$ as a function of $d$.
In this measurement, we observe that ${\rm d}V_n/{\rm d} V_1$ abruptly decreases with increasing $d$.
Therefore there is energy relaxation in this length scale.
This result also shows that we can detect the change of the electronic energy distribution through the change of $V_n$ even in this case without change of the electrochemical potential, although the detail of the mechanism is not perfectly clear at present. ~\cite{comment2}

To evaluate the relaxation length $l_{\rm e}$, we fitted the data with a simple exponential decay.
The solid line in Fig.~\ref{Energy}(b) is the result of the fitting.
The best fitting is obtained for $l_{\rm e}$ of $3\pm 1~\mu$m and this value is much smaller than $l_{\rm t}$.
This result is again consistent with the previous experiments utilizing quantum dots as the local probes, which show the short energy relaxation length 3$~\mu$m.~\cite{2010leSueurPRL}
These two results of the energy relaxation lengths in the cases with electron tunneling and only with energy exchange without tunneling between edge states agree with the previous experiments.
This agreement supports the validity of our measurement method with QPC probes, which can be applied to a wider range of experiments compared to the quantum dot probes.

The possible reason why $l_{\rm e}$ is much smaller than $l_{\rm t}$ is the following.
In the case of relaxation only with energy exchange, the process needs only the interaction to mediate energy transfer between spatially separated edge channels like the long range Coulomb interaction and therefore requires a short relaxation length.
On the other hand in the case of energy relaxation with electron tunneling, the process needs electron tunneling between spatially well separated distant edge channels, which happens rarely and requires a long relaxation distance.
Thus, the energy relaxation with energy exchange will first occur and be followed by relaxation with tunneling of electrons on a longer scale.

\section{Energy Relaxation around Hotspots}

\begin{figure}[b]
\begin{center}
  \includegraphics{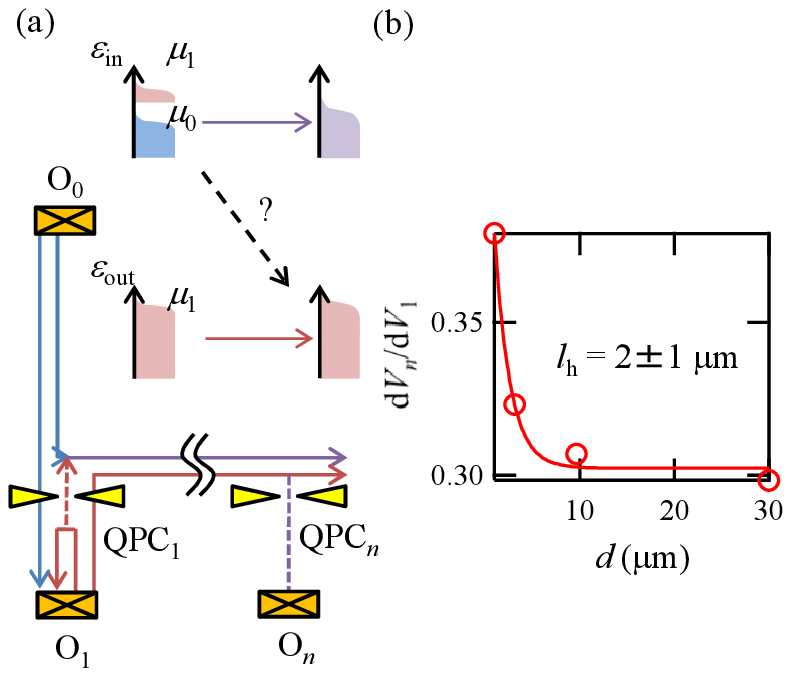}
  \caption{(color online) (a) Schematic of the experimental setup to probe energy relaxation around a hotspot. 
  We set the conductance of QPC$_1$ to $2e^2/h$ and applied high bias across QPC$_1$ to form a hotspot next to QPC$_1$.
  The inner edge channel has the non equilibrium electronic energy distribution formed by the hot electrons supplied by O$_1$ through QPC$_1$.
  (b) ${\rm d}V_n/{\rm d} V_1$ as a function of $d$.
  Decay of ${\rm d}V_n/{\rm d} V_1$ is observed.
  The solid line is the result of the fitting and the evaluated relaxation length $l_{\rm h}$ is $2\pm 1~\mu$m.
  }
  \label{Hotspot}
\end{center}
\end{figure}

Finally, we applied the method with QPCs to measure the local electronic states and the energy relaxation around a hotspot.
One kind of hotspot is known to be formed next to a QPC when we apply a high bias voltage across the QPC.~\cite{1992KomiyamaPRB}
If the applied bias exceeds the potential barrier formed by the voltages of the QPC gates, electrons with high energy pass through the QPC and form hot electrons in the drain.
If we set the conductance of QPC$_1$ to $2e^2/h$ and apply high bias across QPC$_1$, the hot electrons are formed in the inner edge channel.
The electron distribution in the inner channel will be the Fermi distribution originating from O$_0$ plus hot electrons with higher energies supplied by O$_1$ (Fig.~\ref{Hotspot}(a)).
We measured the energy relaxation length around this hotspot with our method utilizing QPCs and discuss the possible mechanism of the energy relaxation from the information of the obtained relaxation length.

Figure~\ref{Hotspot}(b) is the obtained results of ${\rm d}V_n/{\rm d}V_1$ as a function of $d$.
We applied a high bias up to 10~mV on O$_1$, which exceeds the potential barrier formed by the QPC of sub meV, to create a hotspot next to QPC$_1$.
The ${\rm d}V_n/{\rm d}V_1$ abruptly decreases with $d$ in this measurement.
This reflects the increase of the electron temperature in the outer edge channel and there is energy relaxation on this length scale.
The solid line in Fig.~\ref{Hotspot}(b) is the best fit to the data with the relaxation length $l_{\rm h}$ of $2\pm 1\mu $m.
This value is comparable with $l_{\rm e}$ obtained in the case of the energy relaxation only with energy exchange between the edge states (Fig.~\ref{Energy}).
From this result, we can conclude that the possible mechanism of the energy relaxation on this length scale around the hotspot is the exchange of energy between edge states without electron tunneling.~\cite{comment3}

\section{Conclusions}

In conclusion, we have measured the local electronic states and evaluated the energy relaxation length of the quantum Hall edge channels utilizing QPC probes.
We confirmed the relaxation length with electron tunneling as $>30~\mu $m and the relaxation length with energy exchange as $3\pm 1~\mu $m. 
By comparing these results with the previous experiments, we have checked the validity of our detection method with QPCs.
We applied this method to measure energy relaxation around a hotspot and obtained the value of $2\pm 1~\mu$m.
This revealed that the possible relaxation mechanism on this length scale around the hotspot is the energy exchange between edge channels.

\section*{Acknowledgments}

We thank G. Allison, T. Okamoto and T. Obata for fruitful discussions and technical supports.
Part of this work is supported by the Grant-in-Aid for Research Activity start-up and Young Scientists B, Funding Program for World-Leading Innovative R\&D on Science and Technology (FIRST) from the Japan Society for the Promotion of Science, Toyota Physical \& Chemical Research Institute Scholars, RIKEN Incentive Research Project, and IARPA project ``Multi-Qubit Coherent Operations'' through Copenhagen University.


\begin{thebibliography}{9}
\bibitem{1980KlitzingPRL}
K. von Klitzing, G. Dorda, and M. Pepper:
Phys. Rev. Lett. {\bf 45} (1980) 494.

\bibitem{1982HalperinPRB}
B. I. Halperin: Phys. Rev. B {\bf 25} (1982) 2185.

\bibitem{1988ButtikerPRB}
M. B\"uttiker:
Phys. Rev. B {\bf 38} (1988) 9375.

\bibitem{1991BeenakkerSSP}
C. W. J. Beenakker, H. van Houten:
Solid State Phys. {\bf 44} (1991) 1.

\bibitem{2003JiNat}
Y. Ji, Y. Chung, D. Sprinzak, M. Heiblum, D. Mahalu and H. Shtrikman:
Nature {\bf 422} (2003) 415.

\bibitem{2007NederNatPhys}
I. Neder, F. Marquardt, M. Heiblum, D. Mahalu and V. Umansky:
Nature Physics {\bf 3} (2007) 534.

\bibitem{2007FeveSci}
G. Feve, A. Mahe, J.-M. Berroir, T. Kontos, B. Placais, D. C. Glattli, A. Cavanna, B. Etienne and Y. Jin:
Science {\bf 316} (2007) 1169.

\bibitem{2001IonicioiuIJMPB}
R. Ionicioiu, G. Amaratunga and F. Udrea:
Int. J. Mod. Phys. B {\bf 15} (2001) 125.

\bibitem{2004StacePRL}
T. M. Stace, C. H. W. Barnes and G. J. Milburn:
Phys. Rev. Lett. {\bf 93} (2004) 126804.

\bibitem{2010AltimirasNatPhys}
C. Altimiras, H. le Sueur, U. Gennser, A. Cavanna, D. Mailly, and F. Pierre:
Nature Phys. {\bf 6} (2010) 34.

\bibitem{2010OtsukaPRB}
T. Otsuka, E. Abe, Y. Iye and S. Katsumoto:
Phys. Rev. B {\bf 81} (2010) 245302.

\bibitem{2010leSueurPRL}
H. le Sueur, C. Altimiras, U. Gennser, A. Cavanna, D. Mailly and F. Pierre:
Phys. Rev. Lett. {\bf 105} (2010) 056803.

\bibitem{2010AltimirasPRL}
C. Altimiras, H. le Sueur, U. Gennser, A. Cavanna, D. Mailly and F. Pierre:
Phys. Rev. Lett. {\bf 105} (2010) 226804.

\bibitem{2010LundePRB}
A. M. Lunde, S. E. Nigg and M. B\"uttiker:
Phys. Rev. B {\bf 81} (2010) 041311(R).

\bibitem{2012VenkatachalamNatPhys}
V. Venkatachalam, S. Hart, L. Pfeiffer, K. West and A. Yacoby:
Nature Physics {\bf 8} (2012) 676.

\bibitem{1988vanWeesPRL}
B. J. van Wees, H. van Houten, C. W. J. Beenakker, J. G. Williamson, L. P. Kouwenhoven, D. van der Marel, and C. T. Foxon:
Phys. Rev. Lett. {\bf 60} (1988) 848.

\bibitem{1988WharamJPhysC}
D. A. Wharam, T. J. Thornton, R. Newbury, M. Pepper, H. Ahmed, J. E. F. Frost, D. G. Hasko, D. C. Peacock, D. A. Ritchie, and G. A. C. Jones:
J. Phys. C {\bf 21} (1988) L209.

\bibitem{1992KomiyamaPRB}
S. Komiyama, H. Hirai, M. Ohsawa, Y. Matsuda, S. Sasa and T. Fujii:
Phys. Rev. B {\bf 45} (1992) 11085.

\bibitem{1992MullerPRB}
G. M\"uller, D. Weiss, A. V. Khaetskii, K. von Klitzing, S. Koch, H. Nickel, W. Schlapp and R. Losch:
Phys. Rev. B {\bf 45} (1992) 3932.

\bibitem{comment1}
About the absolute values of ${\rm d}V_n/{\rm d} V_1$, those will be affected by the contact resistances of Ohmic contacts.
The real voltages applied on the edge states become smaller than the value applied on the Ohmic contact because of the voltage drop at the Ohmic resistance.
If the value of the Ohmic resistance is negligibly small, ${\rm d}V_n/{\rm d} V_1$ will be close to 1 in the case without energy relaxation.
In this device, however, the value is relatively large and ${\rm d}V_n/{\rm d} V_1$ becomes smaller than 1 even without energy relaxation.

\bibitem{1995HiraiPRB}
H. Hirai, S. Komiyama, S. Fukatsu, T. Osada, Y. Shiraki and H. Toyoshima:
Phys. Rev. B {\bf 52} (1995) 11159.

\bibitem{comment2}
About the possible mechanism of this detection, we think tentatively that the thermoelectric effect might cause the correlation between the electronic energy distribution and the probe voltage $V_n$.
Around the probe contact, there will be an effective temperature gradient between the heated electrons created by the non-equilibrium distribution and the cold electrons in the contact.
This will cause thermoelectric power and we can detect the change of the electron distribution with $V_n$.

\bibitem{comment3}
The relaxation length by electron tunneling will also decrease with application of the bias voltage exceeding the Landau level spacing~\cite{2002WurtzPRB} $7$~mV at 4~T.
But in our measurement, we observed short relaxation length even when the bias is smaller than 7~mV.
So, we think the possible relaxation mechanisms will be the energy exchange between the edge states.

\bibitem{2002WurtzPRB}
A. W\"urtz,, R. Wildfeuer, A. Lorke, E. V. Deviatov, V. T. Dolgopolov:
Phys. Rev. B {\bf 65} (2002) 075303.

\end{thebibliography}
\end{document}